\documentclass{PoS}
\let\ifpdf\relax
\usepackage[authoryear,square]{natbib}
\bibpunct{(}{)}{;}{a}{}{,}
\usepackage{epsfig,graphics}
\usepackage{longtable}
\usepackage{caption}

\newcommand{\hi}{H\,{\sc{i}}\,}
\newcommand{\mhi}{$M_{HI}$\,}
\newcommand{\msol}{M$_{\odot}$}

\newcommand{\eg}{e.\,g.,}
\newcommand{\degr}{^{\circ}}
\newcommand{\HI}{H\,{\sc i}}

\newcommand{\NHI}{$N_{\rm HI}$}

\def\aj{{AJ}}
\def\aa{{A\&A}}
\def\araa{{ARA\&A}}
\def\apj{{ApJ}}
\def\apjl{{ApJ}}

\def\aap{{A\&A}}

\def\mnras{{MNRAS}}

\title{Exploring Neutral Hydrogen and Galaxy Evolution with the SKA}

\ShortTitle{\hi and Galaxy Evolution}

\author{
S.~-L. Blyth$^1$,
J.~M. van der Hulst\footnote{Speaker}\ $^2$,
M.A.W. Verheijen$^2$,
HI SWG Members$^3$,
B. Catinella$^4$,
F. Fraternali$^{5,2}$,
M.~P. Haynes$^6$,
K.~M. Hess$^{1,2,7}$,
B.S. Koribalski$^8$,
C. Lagos$^9$,
M. Meyer$^{10}$,
D. Obreschkow$^{10}$,
A. Popping$^{10,11}$,
C. Power$^{10}$,
L.~Verdes-Montenegro$^{12}$,
M. Zwaan$^9$
\\
$^1$Astrophysics, Cosmology and Gravity Centre (ACGC), Department of Astronomy, University of Cape Town, Private Bag X3, 7701, Rondebosch, South Africa;
$^2$Kapteyn Astronomical Institute, University of Groningen, Postbus 800, 9700 AV, Groningen, The Netherlands;
$^3$SKA HI Science Working Group;
$^4$Centre for Astrophysics \& Supercomputing, Swinburne University of Technology, Hawthorn, VIC 3122, Australia;
$^5$Department of Physics \& Astronomy, University of Bologna, via Berti Pichat 6/2, 40127, Bologna, Italy;
$^6$Center for Radiophysics and Space Research, Space Sciences Building, Cornell University, Ithaca NY, 14853 USA;
$^7$Netherlands Institute for Radio Astronomy (ASTRON), PO Box 2, 7990 AA Dwingeloo, The Netherlands;
$^8$Australia Telescope National Facility, CSIRO Astronomy and Space Science,
P. O. Box 76, Epping, NSW 1710, Australia;
$^9$European Southern Observatory, Karl-Schwarzschild-Strasse 2, D-85748, Garching, Germany;
$^{10}$ICRAR, University of Western Australia, 35 Stirling Highway, Crawley, WA 6009, Australia; 
$^{11}$ARC Centre of Excellence for All-sky Astrophysics (CAASTRO);
$^{12}$Instituto de Astrofisica de Andalucia (IAA/CSIC), Apdo. 3004, 18080, Granada, Spain
\\
E-mail: \email{sarblyth@ast.uct.ac.za}
}

\abstract{
  One of the key science drivers for the development of the SKA is to
  observe the neutral hydrogen, \hi{}, in galaxies as a means to probe
  galaxy evolution across a range of environments over cosmic
  time. Over the past decade, much progress has been made in
  theoretical simulations and observations of \hi{} in
  galaxies. However, recent \hi{} surveys on both single dish radio
  telescopes and interferometers, while providing detailed information
  on global \hi{} properties, the dark matter distribution in
  galaxies, as well as insight into the relationship between star
  formation and the interstellar medium, have been limited to the
  local universe. Ongoing and upcoming \hi{} surveys on SKA pathfinder
  instruments will extend these measurements beyond the local universe
  to intermediate redshifts with long observing programmes. We present
  here an overview of the \hi{} science which will be possible with
  the increased capabilities of the SKA and which will build upon the
  expected increase in knowledge of \hi\ in and around galaxies
  obtained with the SKA pathfinder surveys. With the SKA1 the greatest
  improvement over our current measurements is the capability to image
  galaxies at reasonable linear resolution and good column density
  sensitivity to much higer redshifts ($0.2 < z < 1.7$). So one will
  not only be able to increase the number of detections to study the
  evolution of the \hi{} mass function, but also have the sensitivity
  and resolution to study inflows and outflows to and from galaxies
  and the kinematics of the gas within and around galaxies as a
  function of environment and cosmic time out to previously unexplored
  depths.  The increased sensitivity of SKA2 will allow us to image
  Milky Way-size galaxies out to redshifts of $z=1$ and will provide
  the data required for a comprehensive picture of the \hi{} content
  of galaxies back to $z\sim2$ when the cosmic star formation rate
  density was at its peak. }

\FullConference{
Advancing Astrophysics with the Square Kilometre Array\\
June 8-13, 2014\\
Giardini Naxos, Italy}

\begin{document}

\section{Introduction}

How galaxies form and evolve is one of the fundamental questions in
modern astrophysics. Over the past decade, multiwavelength
observations of galaxies spanning a wide range of cosmic time have
indicated that galaxy evolution seems to depend primarily on two
parameters: a galaxy's stellar mass, and the environment in which it
is located. In the $\Lambda$-cold dark matter (CDM) hierarchical
structure formation picture, galaxies form through the successive
mergers of smaller units (namely dark matter haloes) to form larger
structures. Prior to the epoch of galaxy formation, baryons existed
almost entirely in gaseous form, and it is through the infall of this
material onto the filamentary structures of the cosmic web, its
accretion into the deepest potential wells, and ultimate collapse into
dense molecular clouds, that galaxies were able to start forming and
producing stars. The ongoing influence of dark matter haloes, and
complex feedback mechanisms between baryonic components, continue to
regulate the gas content and evolution of galaxies today.  In order to
fully understand the build up of stellar mass in galaxies, we need to
understand the role that neutral gas, the fuel for star formation,
plays, along with the physical processes involved. Therefore we need
to study the neutral gas distributions and kinematics of galaxies, in
large numbers, in different environments, as a function of cosmic time
to uncover the full picture of galaxy evolution and structure
formation in the Universe. This is one of the main scientific drivers
for the development of the SKA.

Along with recent progress in observations, there have been many
theoretical advances in both semi-analytic modelling and full
hydrodynamical modelling of galaxy evolution and, importantly, the
role of neutral hydrogen gas, in both its atomic (\hi) and molecular
forms (see section~\ref{theory}). However, observations of neutral gas
have lagged behind observations at other wavelengths; due to the
intrinsic faintness of \hi emission, unreasonably long observing times
are required to probe to redshifts beyond $z\sim 0.25$~ with existing
facilities, although with the construction of SKA pathfinder
instruments and significant upgrades to current facilities, deeper,
more sensitive, observations will be possible in the very near future
(see section~\ref{SKApathfinders}). The SKA will allow us to probe HI
emission in galaxies to previously unexplored depths and further back
in cosmic history. The SKA2 will enable optical quality \hi\ imaging
of Milky Way size galaxies out to redshifts of $z=1$
 
Analysis of thousands of galaxies in the Sloan Digital Sky Survey
(SDSS) has shown that they follow a bimodal distribution in colour as
a function of stellar mass~(\cite{2004ApJ...600..681B}); most galaxies
are either located in the so-called blue cloud or on the red sequence,
indicating that star formation is either still ongoing or was quenched
billions of years ago. Furthermore, the galaxy `main sequence' (star
formation rate vs. stellar mass) shows a relatively smooth evolution
over cosmic time from $z\sim2.5$ to $z\sim 0$ while over the same
period, the star formation rate density of the Universe has dropped by
more than an order of magnitude since its peak at $z \sim
2-3$~(\cite{1998ApJ...498..106M, 2006ApJ...651..142H}). Star formation
also seems to have shifted from the more massive galaxies to less
massive galaxies at later times, a pattern known as cosmic
downsizing~(\cite{1996AJ....112..839C}). Understanding the role of the
neutral gas that fuels star formation and enables the build up of
stellar mass as well as the role of the environment in enabling or
quenching these processes, is vital to disentangle the physics
involved in the evolution of galaxies. To move forward, a number of
key questions related to the role of \hi in galaxies need to be
answered such as:

\begin{itemize}
\item [-] \vspace{-1pt} What is the distribution and kinematics of the
  neutral (\hi) gas within and around galaxies, both as a function of
  environment (i.e. groups/clusters vs. the field) and over cosmic
  time?
\item [-] \vspace{-5pt} How much \hi\ is there on average as a
  function of redshift?
\item [-] \vspace{-5pt} How does the \mhi of galaxies scale with their
  stellar/halo masses and other properties, \eg star formation rate,
  as a function of environment and redshift?
\item [-] \vspace{-5pt} How important is gas accretion vs. merging in
  terms of building stellar mass?
\end{itemize}

The process of star formation is intimately related to how individual
galaxies evolve. On kpc-size scales, the Kennicutt-Schmidt
law~(\cite{2008AJ....136.2846B,2012AJ....144....3L,2012ApJ...752...98C})
describes the relationship between the star formation rate surface
density and the molecular gas mass surface density. However, details
of the conditions needed for star formation to begin at the scales of
(molecular) clouds are as yet unclear. High resolution observations of
the interstellar medium (ISM) in individual galaxies are required in
order to clarify the astrophysical processes taking place at these
scales. This topic will be more fully discussed in a separate
chapter,~{\it The Interstellar medium in Galaxies}
(\cite{deBlok2014}). Further discussion of the interaction of galaxies
with their environment, the IGM and the distribution of neutral
hydrogen in the cosmic web will be presented in the chapter~{\it The
  Intergalactic Medium and the Cosmic Web} (\cite{Popping2014}).

Observations of \hi in absorption along lines of sight to strong radio
sources can provide information on the neutral gas content of galaxies
at higher redshifts than it is possible to reach directly with \hi
emission measurements. These are key measurements for studies of
galaxy evolution at higher redshifts, and will be discussed further in
a separate chapter,~{\it Cool Outflows and HI absorbers}
(\cite{Morganti2014})

\section{\hi observations and theory: the status quo}

\subsection{Recent observational results}

\subsubsection{The Local Universe}

Recent years have ushered in a ``golden age'' of \hi\ science that has
seen a variety of \hi\ surveys to study galaxy evolution take
place. The recent single dish, large area blind \hi\ surveys HIPASS
(\cite{2001MNRAS.322..486B}) and ALFALFA (\cite{2005AJ....130.2598G})
have shed light on the global \hi\ properties of galaxies in general
and provided the best estimates of the global \hi\ mass function. Poor
statistics at the very high \hi\ mass end {\bf and} the very low \hi\
mass end ($< 10^7$ \msol) limit the current results and do not resolve
questions such as whether the \hi\ mass function depends on the local
enviroment (\cite{2005MNRAS.359L..30Z,2005ApJ...621..215S}). Yet these
surveys have revealed that there is indeed an extreme paucity of \hi\
objects without stellar counterparts when the \hi\ mass detection
limit is pushed well below $5 \times 10^6$ \msol, enabling the
interesting discovery of a few local, gas-rich dwarf galaxies such as
Leo T (\cite{2008MNRAS.384..535R}) and Leo P
(\cite{2013AJ....146...15G}). In addition the surveys demonstrate the
power of combining the \hi\ survey information with data from other
wavebands (optical, UV, IR) to address questions relevant to galaxy
evolution
(\cite{2010MNRAS.403..683C,2010MNRAS.408..919S,2009MNRAS.399.2264W}).

More detailed information about the kinematics and structure of the
\hi\ disks in individual galaxies has come from resolved observations
using synthesis radio telescopes.  Major surveys carried out in the
last decade are WHISP (\cite{2001ASPC..240..451V}), THINGS
(\cite{2008AJ....136.2563W}), LVHIS (\cite{2008glv..book...41K}),
Atlas-3D (\cite{2012MNRAS.422.1835S}), Little THINGS
(\cite{2012AJ....144..134H}), VLA-ANGST (\cite{2012AJ....144..123O}),
VIVA (\cite{2009AJ....138.1741C}), SHIELD
(\cite{2011ApJ...739L..22C}), FIGGS (\cite{2008MNRAS.386.1667B}),
HALOGAS (\cite{2011A&A...526A.118H}) and BlueDisks
(\cite{2013MNRAS.433..270W}). Major results are the systematic
inventory of the distribution of dark matter in galaxies, the
prevalence of warps, tidal features, and signs of ram pressure
stripping, and detailed relationships between star formation and the
ISM including galactic fountains and the presence of extra-planar
\hi{}. A synopsis of results to date can be found in two recent
reviews (\cite{2008A&ARv..15..189S,2013pss6.book..183V}).

Observations of \hi{} in galaxies in very dense environments (a well
studied example is the Virgo cluster (\cite{2009AJ....138.1741C})
demonstrate the importance of the environment.  The growing inventory
of detailed \hi{} observations of galaxies in different environments
is mixed, but suggest that even intermediate density environments in
galaxy groups (\cite{2013AJ....146..124H}), and on the outskirts of
clusters (\cite{2013MNRAS.431.2111J}) impact the HI content of
galaxies.  Dense environments in small volumes, such as Hickson
compact groups, also exhibit \hi{} deficiencies
(\cite{2001A&A...377..812V}) with a suggestion that a reservoir of low
density \hi{} might exist to feed the galaxies
(\cite{2011ApJ...729..149B}). In very isolated environments, such as
voids (\cite{2012AJ....144...16K,2013AJ....145..120B}) there is a
suggestion that accretion is an important process. Obtaining
information for galaxies that are specifically selected to be
extremely isolated will be very important (\cite{2011A&A...532A.117E})
provided detailed imaging is performed
(\cite{2014A&A...567A..56S}). Having a good enough sample of \hi{} in
and around many galaxies in different enviroments to address the
question of {\it nature versus nurture} will will be a prime goal for
the SKA.

Three surveys are currently pushing the present technology to
redshifts beyond $z=0.2$; the HIGHz survey
(\cite{2008ApJ...685L..13C}) with Arecibo and two interferometric
surveys, BUDHIES (\cite{2007ApJ...668L...9V,2013MNRAS.431.2111J}) and
CHILES (\cite{2013ApJ...770L..29F}) which are using observing times
approaching 1000 hours.  To date, the most successful method of
detecting \hi{} emission beyond $z=0.2$ with current instrumentation
is stacking based on position and redshift information from optical
surveys. In the following sections we will briefly describe the
current status of \hi\ observations and then focus on the important
next steps to be addressed by SKA1 and SKA2.

\subsubsection{Observations at low redshift ($z<0.4$)}

Exploiting the exquisite sensitivity of the Arecibo radio telescope,
the HIGHz survey provides a glimpse into the \hi{} properties of
star-forming galaxies at $z\sim 0.2$ located in relatively isolated
environments.  The survey measured the \hi{} content of 39
optically-selected galaxies with redshifts between $0.17 < z < 0.25$;
these are all actively star-forming, disk-dominated systems with
stellar and \hi{} masses larger than $10^{10}$ M$_\odot$. This sample
includes not only the highest-redshift detections of \hi{} emission
from individual galaxies to date, but also some of the most
\hi{}-massive systems known. Despite being exceptionally large, the
\hi{} reservoirs of these galaxies are consistent with what is
expected from their UV and optical properties.This, and the fact that
the galaxies lie on the baryonic Tully-Fisher relation, suggests that
HIGHz systems are rare, scaled-up versions of gas-rich disks in the
local Universe. These observations provide important insight into the
properties of the massive, \hi{}-rich systems that will likely
dominate the next-generation \hi{} surveys with the SKA and its
pathfinder telescopes.

Taking advantage of the upgraded Westerbork receiver system, a Blind,
Ultra-Deep \HI\ Environmental Survey (BUDHIES) has been carried out to
image two cosmic volumes, both covering the redshift range
z=0.164$-$0.224, centered on the galaxy clusters Abell 2192 at
$z=0.187$ and Abell 963 at $z=0.206$. A963 is a massive, lensing,
X-ray bright, Butcher-Oemler cluster with an unusual high fraction of
blue galaxies in its center. The main scientific goal of BUDHIES is to
characterize the gas content of these blue galaxies in relation to the
gas content of galaxies in the cluster outskirts and the surrounding
field in which this cluster is embedded. A~2192 is a less massive
cluster that mainly serves as a control object. The clusters within
their Abell radii occupy only $\sim$5\% of the total surveyed volume,
which also contains voids and galaxy overdensities in the fore- and
background of both clusters.

Data were collected with the WSRT with 78$\times$12$^{\rm hr}$ on
A2192 and 117$\times$12$^{\rm hr}$ on A963 to obtain a similar HI mass
limit of 2$\times$10$^9$ M$_\odot$.  A total of 39 galaxies were
detected in the volume containing A2192 and 120 galaxies in
A963. First results from a pilot survey showed that none of the blue
galaxies in A963 were detected, while stacking the HI spectra from
galaxies of similar luminosity and color in the surrounding field
showed a clear detection (\cite{2007ApJ...668L...9V}). The full
dataset is currently being analysed with a focus on the \HI\ content
of galaxies in the various substructures within and around the
clusters (\cite{2012ApJ...756L..28J, 2013MNRAS.431.2111J}), as well as
the \HI\--based Tully-Fisher relation at these redshifts.

The COSMOS \hi{} Large Extragalactic Survey (CHILES) is an ongoing
1000 hour program designed to take advantage of the upgraded
capabilities on the Karl G. Jansky Very Large Array (VLA). The upgrade
enables continuous redshift coverage for \hi{} in emission from
$z=0-0.5$, doubling the look-back time of existing \hi{}
measurements. A single pointing centered in the COSMOS field, CHILES
is expected to detect and resolve Milky Way-like galaxies in \hi{}
($M_{HI}=3\times10^{10}$~M$_{\odot}$) at the most distant redshift
accessible, and the most gas-rich galaxies that we know of at $z=0$,
totalling over 300 \hi{}-rich objects in the roughly $4 \times 10^5$
Mpc$^3$ surveyed.  The chosen field intersects with known large-scale
structure at low, intermediate, and high redshifts, and COSMOS
provides a wealth of complementary multi-wavelength photometry and
optical spectroscopy enabling the study of galaxy evolution as a
function of galaxy properties, environment, and cosmic time.  CHILES
will be the first survey to probe the \hi{} content, growth, and
morphology of galaxies over the same redshift range in which star
formation undergoes ``cosmic downsizing''.

First results from 50 hours observing (\cite{2013ApJ...770L..29F})
produced 33 direct detections in observations covering a continuous
redshift range $z=0-0.193$ over a $34'\times34'$ field of view,
including three galaxies for which there were no previous
spectroscopic redshifts. The most distant galaxy was detected at
z=0.176 and has an HI mass of $8\times10^{9}$~M$_{\odot}$.  Stacking
the \hi{} spectra of 80 galaxies in a ``wall'' between $z=0.12-0.13$
produced a detection with an average HI mass of $1.8\times10^9$
M$_{\odot}$.


\smallskip
\noindent \textbf{Recent results from \hi stacking} \\
\hi\ observations have only begun to detect 21 cm line emission from
individual galaxies beyond redshift $z\sim 0.1$.  Although a number of
galaxies will be individually detected at these and higher redshifts
by the upcoming SKA pathfinder surveys (see Table~\ref{tabsurveys}),
the most stringent constraints on

\begin{minipage}{0.5\textwidth}
	\hspace{-25pt}
	\includegraphics[width=8.5cm]{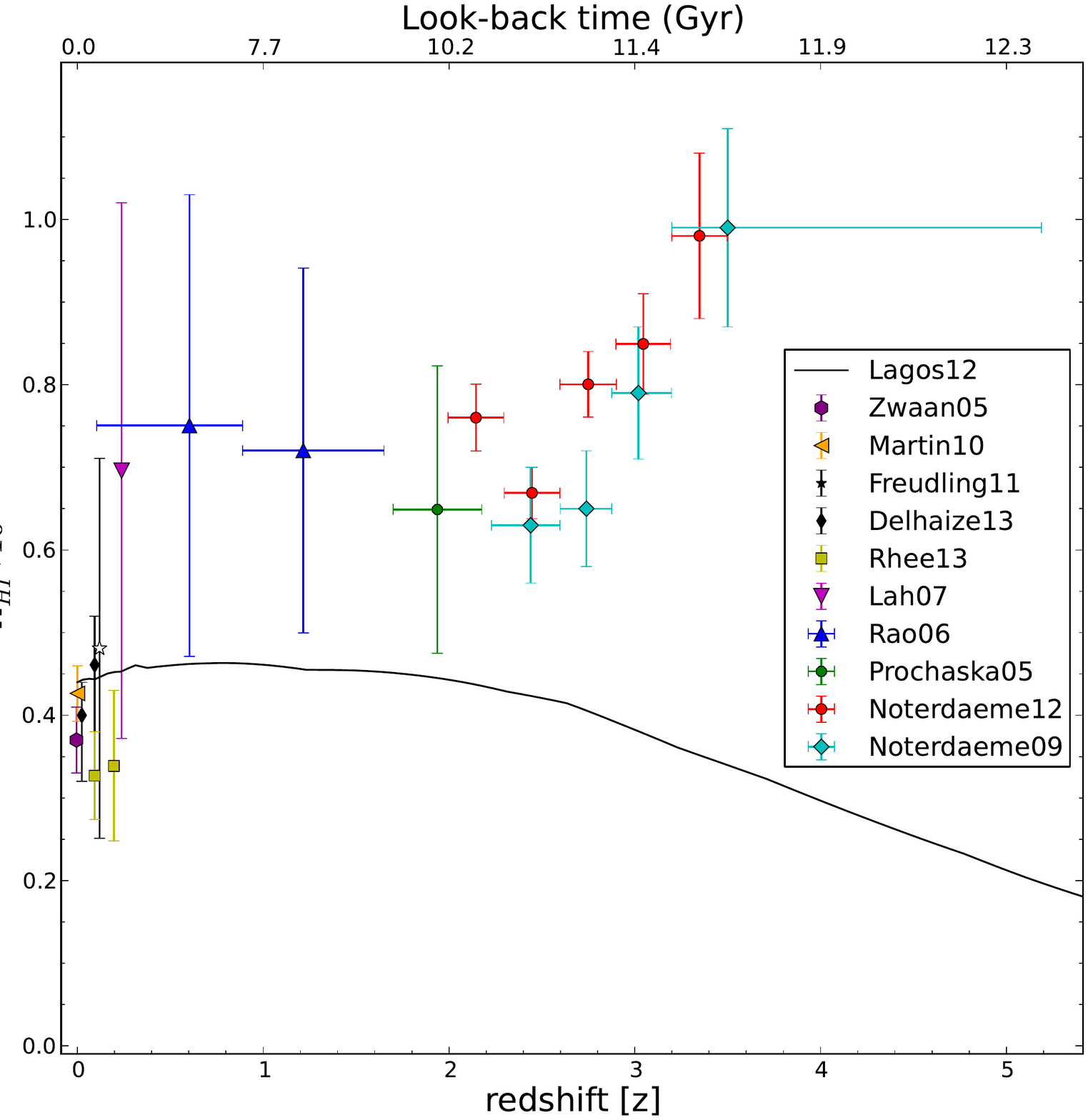}
\label{OmegaHI}
\end{minipage}	
\begin{minipage}{0.45\textwidth}
\center
\captionof{figure}{\small
  Recent compilation of cosmic \hi{} gas density measurements as
  functions of redshift and look-back time. At z=0, the hexagon and
  triangle refer to direct observations from blind \hi surveys
  (\cite{2005MNRAS.359L..30Z,Martin10}). All points above $z=0.4$ are
  damped Ly-$\alpha$ measurements from HST and SDSS
  (\cite{2005ApJ...635..123P,2006ApJ...636..610R,2009A&A...505.1087N,2012A&A...547L...1N}). Bridging
  the gap between the two are the estimates from \hi{} stacking of
  Parkes (diamonds)(\cite{2013MNRAS.433.1398D}, WSRT (yellow
  squares)~(\cite{2013MNRAS.435.2693R}) and GMRT (magenta
  triangle)~(\cite{2007MNRAS.376.1357L}) observations. The curve
  represents model predictions by \cite{2012MNRAS.426.2142L}.}
\end{minipage}	

\noindent
the \hi\ content of galaxies across cosmic time will come from
stacking of undetected sources.

Stacking has become a common tool to constrain the statistical
properties of a population of objects that lack individual detections
in a survey, and has been applied to a variety of different
astrophysical data, including \hi-line spectra
(\cite{2001A&A...372..768C,2007MNRAS.376.1357L,2007ApJ...668L...9V,2009MNRAS.399.1447L,2011MNRAS.411..993F,2013MNRAS.433.1398D,2013A&A...558A..54G}).
This technique requires independent measurements of the redshifts of
the galaxies, and yields an estimate of the average \hi\ content of a
sample of galaxies by co-adding their line emission (see description
in \cite{2011MNRAS.411..993F}).

Over the past decade, in an effort to probe the evolution of \hi\ gas
with cosmic time, Lah et al. (2007), who used the Giant Metrewave
Radio Telescope (GMRT) to estimate the \hi\ content of star-forming
galaxies at $z\sim 0.24$, and Rhee et al. (2013), who targeted field
galaxies at redshift $z\sim 0.1$ and $z\sim 0.2$ with the Westerbork
Synthesis Radio Telescope (WSRT), showed that it is possible to
constrain the cosmic \hi\ gas density, $\Omega_{HI}$, of the Universe
up to look-back times 2-4 Gyr with \hi\ stacking. Intriguingly, these
studies have begun to bridge the gap between high-z damped
Lyman-$\alpha$ observations and blind 21-cm surveys at $z=0$
(Fig. 1). 
Determining $\Omega_{HI}$ is especially critical at these intermediate
redshifts when the cosmic star formation rate density plummeted.

Stacking is also extremely promising for galaxy evolution studies in
general, as a means to investigate the connection between gas, star
formation rate, and other galaxy properties across environment and
time.  Application of the stacking technique to a sample of $\sim$5000
optically-selected galaxies with \hi\ observations from ALFALFA has
successfully recovered the main scaling relations connecting the gas
fractions of galaxies with their structural and star formation
properties~(\cite{2011MNRAS.411..993F}) and the effects of environment
on gas content~(\cite{2012MNRAS.427.2841F}).  Because of the large
statistics needed for these studies, and because of the need to reach
the gas-poor regime, this question is optimally addressed by \hi\
stacking, rather than by individually detected galaxies. Environmental
effects on the gas have also been investigated at much greater
distances, using \hi\ stacking at $z\sim
0.2$~(\cite{2007ApJ...668L...9V}) and
$z=0.37$~(\cite{2009MNRAS.399.1447L}).

Clearly, the upcoming \hi\ surveys with the SKA and its pathfinders will allow us to extend all these studies to significantly higher redshifts, and to vastly larger optically-selected samples than is possible today. \hi\ stacking will be a very powerful technique to exploit the next-generation \hi\ surveys, well beyond the limits imposed by individual detections.\\

\subsection{Recent theoretical advances}\label{theory}

The two most widely used techniques to study galaxy formation in a
cosmological context are hydrodynamical simulations and
semi-analytical models (SAMs). Hydrodynamical simulations follow the
evolution of both the dark matter and baryons (gas and stars)
simultaneously, while SAMs evolve the baryons in a pre-computed
(usually by an $N$-body simulation) dark matter background. These
approaches are complementary: both approximate the key physical
processes that are fundamental to galaxy formation, such as star
formation and feedback, but hydrodynamical simulations directly solve
the equations of gas dynamics and the gravitational interactions
between the gas, stars and dark matter, whereas SAMs must approximate
them. SAMs are computationally inexpensive, can model galaxy
populations in cosmological volumes over cosmic time, and can be used
to explore the parameter space of galaxy formation rapidly. In
contrast, hydrodynamical simulations are computationally expensive but
can capture the detailed spatial, kinematical and dynamical structure
of gas, stars, and dark matter in galaxies within individual systems
or in small volumes, and can be used to inform the choice of
physically motivated recipes used in SAMs.  Both approaches have shown
substantial progress in the treatment of gas phases in and outside
galaxies in the last $\sim$5 years.

Hydrodynamical simulations have witnessed important advances in the
treatment of the interstellar medium (ISM) in galaxies and in metal
enrichment of the circumgalactic medium (CGM).  \emph{Non-cosmological
  simulations} of galaxy formation have been used to study in detail
processes such as H$_2$ formation and destruction, the \hi{} to H$_2$
transition in non-equilibrium chemistry, non-equilibrium thermal
state, and variations in the strength of the radiation field
(e.g. Pelupessy et al. 2006, 2009; (\cite{Robertson08, Gnedin09,
  Dobbs11, Shetty12, Glover12, Christensen12}).  This work has allowed
for physical interpretation of the observed relation between the SFR
and the H$_2$ mass surface densities observed in local
galaxies. Another important area of progress has been in exploration
of the mechanisms that drive turbulence in the ISM. (\cite{Dobbs11}
simulated a large spiral galaxy and showed that stellar feedback is
important in setting the scale height of the galaxy disk through its
effect on the velocity dispersion. Similarly, \cite{Shetty12} find
that supernovae are the primary source of turbulence in the dense
environments of starburst galaxies.  However, \cite{Bournaud10} found
in their simulation of a LMC-like dwarf galaxy that gravitational
instabilities -- which drive the velocity dispersion in their
simulation -- rather than stellar feedback set the scale height, while
stellar feedback is necessary to maintain large-scale turbulence,
which initiates cascades to small scales and suppresses the formation
of very dense, small gas clumps (see also \cite{Hopkins12b} and
\cite{Shetty12}). In \emph{cosmological hydrodynamical simulations},
the transition from \hi{} to H$_2$ has an important effect on the
predicted number of large column density absorbers, $N_{\rm H}>7\times
10^{21}\, {\rm cm}^{-2}$ (\cite{Altay10}). Extensive testing of the
relation between galaxies and \hi{} absorbers at high redshifts has
been presented by \cite{Rahmati14}, who reach conclusions similar to
SAMs (\cite{Lagos11b,Lagos14,Popping13,2012MNRAS.424.2701F}), that a
large fraction of the damped Ly-$\alpha$ absorbers are associated with
galaxies with stellar masses $M_{\star}<10^9\,M_{\odot}$ (see also
\cite{Dave13,Bird14,2014arXiv1408.2531R}).

In \emph{semi-analytical models} of galaxy formation, significant
effort has focused on improving the modelling of the ISM in
galaxies. Whereas previous generations of models followed the
evolution of only a single cold gas phase (\cite{2009ApJ...698.1467O,
  Power10}), current models partition this cold gas phase into \hi{}
and H$_2$ phases and follow the evolution of these phases in a
self-consistent manner (e.g. \cite{Cook10, Fu10, Lagos11a,
  Popping13}). These models assume gas in hydrostatic and chemical
equilibrium and use either the empirical relation of (\cite{Blitz06}),
which relates the H$_2$/\hi{} gas surface density ratio to the local
hydrostatic pressure, or theoretical relations that calculate H$_2$
abundance as a result of the equilibrium between rates of destruction
due to UV radiation and formation on dust grains
(e.g. (\cite{Krumholz09, Gnedin11}). Models based on the empirical
relation of Blitz \& Rosolowsky appear to provide a better fit to the
$z$=0 \hi{}\ mass function (\cite{2005MNRAS.359L..30Z, Martin10}),
with good agreement extending down to \hi{}\ masses of $10^6\,
M_{\odot}\, h^{-2}$ (\cite{Lagos11b}, and also reproduce the observed
clustering of \hi{}\ selected galaxies (\cite{Kim12}).

One of the key results of SAMs is shown in Fig.~$1$, which shows the
very mild evolution of $\Omega_{HI}$ from $z=2$ to $z=0$. This
prediction relates to a key question in extragalactic astrophysics:
what accounts for the dramatically different evolution of the star
formation rate (SFR) and \hi{} densities of the
Universe. \cite{Lagos11b} and \cite{Lagos14} argue that the steep
decline observed in the SFR density with decreasing redshift is
closely connected to the steep decline of the molecular gas density --
because star formation is linked explicitly to the H$_2$ density in
galaxies -- whereas the \hi{}\ density is predicted to evolve very
weakly with redshift. Lagos et al. explain this trend as arising from
a combination of decreasing gas fractions and increasing galaxy sizes
with decreasing redshift, both of which act to reduce gas surface
density and the hydrostatic pressure of the disk, which controls the
rate at which stars form. Therefore, the SFR density evolution
reflects the evolution of the surface density of gas in the galaxies
that dominate the SFR in the universe as a function of cosmic time
(see also \cite{Obreschkow09}). This interpretation also implies an
increase in the molecular to dynamical mass ratios and H$_2$/\hi{}
mass ratio with increasing redshift, which is consistent with
observations (\cite{Geach11, Tacconi13}).
 
Recent progress in the simulation and modelling of galaxy formation
has been rapid, and has emphasized the critical role \hi{}\ plays in
driving galaxy formation and evolution. This work has also highlighted
how \emph{pivotal \hi{}\ surveys on the SKA and its pathfinders will
  be} in revealing the full extent of this role, how they will make a
crucial contribution to multi-wavelength surveys campaigns, and how
they will provide -- arguably -- the most powerful tests of the
predictions of our theories of galaxy formation and evolution. For
example, simulations make clear predictions


\begin{longtable}{|l|} \hline 
\textbf{BLIND \hi{} SURVEYS: WIDE-AREA vs. DEEP }  \\
\textbf{Main aims:} To investigate: \\
- Evolution of \hi{} content (\hi{} mass function) of galaxies vs. $z$ and environment \\
- Galaxy formation \& missing satellite problem in Local Group \\*
- Correlation of \hi{} properties of galaxies and with stellar/halo masses,  star-formation rates, etc. \\
- Physics governing cool gas distribution \& evolution at low $z$ \\
- Baryonic Tully-Fisher relation vs. $z$ \\
- Cosmological parameters relating to gas-rich galaxies \\
- Cosmic web \\

\hline \hline
	\begin{tabular}{p{5cm}|p{10cm}}
		CHILES (JVLA) \ \ \ \ \ [\textit{ongoing}]	&  \textbf{Specifications:} Single pointing in COSMOS field, for $z<0.45$ 
        \\		
	\end{tabular} \\ \hline	

	\begin{tabular}{p{5cm}|p{10cm}}
	Medium-deep blind~imaging (APERTIF)	\ \ \ \ \ \ \ \ \ \ \ [\textit{proposed}]				&   \textbf{Specifications:} $z < 0.25$ over 500 deg$^2$
    \\ 									
	\end{tabular} \\ \hline	
	
	\begin{tabular}{p{5cm}|p{10cm}}
		Northern sky \hi{} shallow	\ \
		survey	(APERTIF) \ \ [\textit{proposed}]
        &  \textbf{Specifications:} \ \ $\delta > +27\degr$, $0<z<0.26$ 
	\end{tabular} \\ \hline	
		
	\begin{tabular}{p{5cm}|p{10cm}}
		WALLABY~(ASKAP) \  \  \  \  \  \  \  \  \  \  \  \  \  [\textit{approved}]		 		& \textbf{Specifications:}  $-90\degr < \delta < +30\degr$, $z<0.26$ 
        \\  
											
	\end{tabular} \\ \hline		
										    			
	\begin{tabular}{p{5cm}|p{10cm}}
		DINGO (ASKAP) \ \ [\textit{approved}]  
										& \textbf{Specifications:}  2 Phases: \\
										& \begin{tabular}{l}
										DEEP: $z<0.26$ over 150 deg$^2$ \\
										UDEEP: $0.1~<~ z~< ~0.43$ over 60 deg$^2$ 
										\end{tabular} \\
	\end{tabular} \\ \hline			  

	\begin{tabular}{p{5cm}|p{10.cm}}
		LADUMA~(MeerKAT) \  \  \  \  \  \  \  \  \  \  \  \  [\textit{approved}]	& \textbf{Specifications:}  Single pointing encompassing ECDF-S, $z~<~1.4$ over $\sim$4 deg$^2$, 
        \\  
											
	\end{tabular} \\ \hline	 \hline
											    	
\textbf{ \hi{} SURVEYS OF INDIVIDUAL GALAXIES / CLUSTERS}  \\ 
\textbf{Main aims:} To investigate: \\
- The range of conditions in ISM in local galaxies \\
- The gas-cycle in nearby galaxies by probing high vs. low column density regions \\
- The connection to local cosmic web\\
- Gas-stripping processes in clusters\\ 
- \hi{} content vs. galaxy morphological type \\
\hline \hline		
	\begin{tabular}{p{5cm}|p{10cm}}
		MHONGOOSE	(MeerKAT) [\textit{approved}]			 		& \textbf{Specifications:}  30 nearby galaxies with $10^5 < M_{HI} <10^{10} M_{\odot}$,  $N_{HI}(5\sigma)$ $\sim 10^{19}$ cm$^{-2}$ 
		\\
	\end{tabular} \\ \hline		

	\begin{tabular}{p{5cm}|p{10.cm}}
		Fornax \hi{} survey (MeerKAT)	[\textit{approved}]	 		& \textbf{Specifications:}  Fornax cluster galaxies over 11 deg$^2$ down to $M_{HI}(3\sigma)\sim~5\times~10^5 M_{\odot}$, $N_{HI}(5\sigma)$ $\sim 10^{19}$ cm$^{-2}$ \\
	\end{tabular} \\ \hline	

\caption{\small Overview of upcoming \hi{} emission line surveys on SKA pathfinder facilities, APERTIF [Medium-deep blind imaging survey, Northern sky \hi{} shallow survey (\cite{2009pra..confE..10V}) ], ASKAP [WALLABY and DINGO] (\cite{2012MNRAS.426.3385D}) and MeerKAT [LADUMA (\cite{2012IAUS..284..496H}), MHONGOOSE (mhongoose.astron.nl) and the Fornax \hi{} survey (\cite{2011fvc..confP..49S})].}
\label{tabsurveys}
\end{longtable}


\noindent
for the expected relation between galaxy properties (such as stellar
mass and SFRs), environment (measured by halo mass and also
clustering) and \hi{}\ content (Lagos et al. 2011, 2014, \cite{Dave13,
  Kim13b, Rahmati14}), as well as the clustering of \hi{}\ selected
galaxies at different \hi{}\ masses and cosmic times. These
measurements will be partially accessible with the ASKAP and MeerKAT
pathfinders, but it will be with SKA1 that the most complete and
powerful tests will be made.

\subsection{Upcoming SKA pathfinder \hi\ surveys}\label{SKApathfinders}

Various \hi\ surveys are planned for the SKA pathfinder instruments
MeerKAT, ASKAP and APERTIF over the next 5--7 years, ranging from deep
observations of particular galaxies or clusters to very wide-area,
relatively shallow surveys to narrow, deep surveys. A summary of the
science aims and specifications for those surveys focusing on galaxy
evolution is presented in Table~\ref{tabsurveys}.

This table is meant to be a summary of \hi\ line work to be carried
out in the next 5--10 years which will form a solid basis for further
work (more detailed and sensitive studies of selected objects and
areas and pushing knowledge of \hi\ in and around galaxies to higher
redshifts with the increased collecting area of SKA1 and eventually
the SKA2). The SKA2 will not only be capable of providing optical
quality ($\sim 1"$ resolution) \hi\ images of Milky Way size galaxies
at $z=1$, but will also have the survey speed to carry out the billion
galaxy survey that will chart out the large scale structure of the
Universe from $z=0$ to $z=3$ and allow precision cosmology studies
(\cite{2009astro2010S.219M}).

\section{Probing galaxy evolution with the SKA}

Once SKA1 comes online, the various surveys mentioned above will have
pushed our knowledge of \hi\ in galaxies and its role in galaxy
evolution to a situation where we have a good inventory of detailed
knowledge of the nearby universe.  Shallow surveys will have covered
most of the sky out to $z \sim 0.25$ (\cite{2012MNRAS.426.3385D}) and
provided \hi\ masses of half a million galaxies, of which several tens
of thousands of galaxies will be reasonably well resolved. Deeper
surveys will have pushed to $z \sim 0.6$, but lack the detailed
information needed to characterize the physical processes that govern
galaxy evolution from the \hi\ point of view: gas accretion, gas
removal by winds, ram pressure or tidal stripping. Such details are
also an important ingredient for comparisons with and input to the
advanced modelling described in section 2.2. The obvious next steps
are to obtain more sensitive and more detailed observations in the
nearby universe ($0 < z < 0.25$) for characterizing these processes
and to push single-galaxy \hi\ mass detections to larger distances and
lower \hi\ masses to explore the dependence on cosmic time and
environment of the \hi\ mass function and other scaling relations. In
the following subsections, we briefly describe these science goals in
more detail.

\subsection{The \hi\ mass function} 

The \HI\ mass function (\cite{1993ApJ...419..515R}) is one of the most
important statistical measures for gas-rich galaxy populations. It
describes the volume density of galaxies as a function of \HI\ mass
and is therefore the neutral hydrogen equivalent of the optical
luminosity function. One important use of the \HI\ mass function in
the context of galaxy evolution is its second moment, which gives the
cosmic \HI\ mass density: the number of neutral hydrogen atoms per
unit volume. This is an important input into the understanding of the
evolution of the cosmic star formation rate density. When combined
with damped Ly\,$\alpha$ measurements at higher redshifts (e.g.,
Noterdaeme et al. 2012, and references therein), the evolution of the
\HI\ mass density can be charted and compared to models of gas and
galaxy evolution. The \HI\ mass density can be compared with the
predictions of semi-analytic or hydrodynamic models in order to study
the processes governing the distribution and evolution of cool gas
(e.g., \cite{Obreschkow09}, Lagos et al. 2011).

Of equal importance is the {\em shape} of the \HI\ mass function. For
example, Kim et al. (2012) demonstrate that the \HI\ mass function is
a more sensitive probe of cosmological reionization for galaxy
formation than the optical galaxy luminosity function. For
semi-analytic models, the \HI\ mass function shape at $z=0$ is an
indispensable constraint (\cite{Obreschkow09}, Lagos et al. 2010,
Neistein et al. 2010, Kim et al. 2011). The \HI\ mass function puts
additional limits on the models compared to, e.g., the stellar mass
function because the correlation between halo mass and cold gas mass
is very different from that between halo mass and stellar mass (the
most gas-rich galaxies often have low halo mass). Furthermore, Kim et
al. (2012) show how different star formation laws affect the shape of
the \HI\ mass function.

\begin{figure}[!t]
\centering
\includegraphics[width=\textwidth]{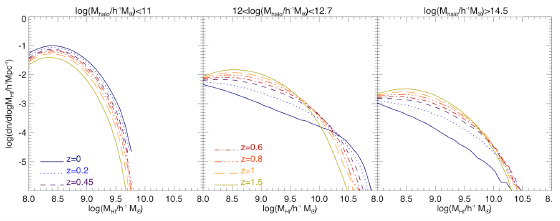}
\caption{\small
Models from Lagos et al. (2012) showing evolution of the \hi{} mass function with redshift and halo mass. Halo mass bins correspond to field (left panel), group (middle panel), and cluster (right panel) environments.}
\label{figHIMF}
\end{figure}

 The \HI\ mass function has been well measured for the Local Universe
(\cite{2005MNRAS.359L..30Z, Martin10}; however, controversy still
surrounds the magnitude and sign of the environmental density
dependence of the \HI\ mass function, possibly due to depth and cosmic
variance issues with existing shallow surveys
(\cite{2005MNRAS.359L..30Z, 2005ApJ...621..215S}). Upcoming precusor
surveys such as WALLABY, will detect galaxies over a $\sim$100 times
larger volume and will therefore be much less sensitive to cosmic
variance, enabling detailed study of the variation of the \HI\ mass
function with environmental density. The deeper HI pathfinder surveys
such as DINGO and LADUMA will enable measurement of the \HI\ mass
function to intermediate redshifts, $z<0.6$.

With the SKA, it will be possible to push measurements of the \HI\
mass function to even higher redshifts ($z\sim1$), which will in turn
place even more stringent constraints on galaxy evolution models (see
Fig.~\ref{figHIMF}). It will also allow us to probe the evolution of
the cosmic neutral gas density, $\Omega_{gas}$, from \HI\ emission
into the redshift range where it is currently estimated using damped
Ly-$\alpha$ and magnesium absorber measurements
(\cite{2005ApJ...635..123P, 2006ApJ...636..610R, 2009A&A...505.1087N,
  2012A&A...547L...1N, 2013A&A...556A.141Z}).

Many questions demanding measurements of the \HI\ masses of galaxies can also be investigated with these observations. As functions of both environment (i.e., field vs group or cluster)(e.g., \cite{2013AJ....146..124H, 2013MNRAS.431.2111J}) and cosmic time, we will be able to study, for lower \HI\ masses than ever before, the relationships between \HI\ mass and stellar mass, star formation rate, galaxy morphology, etc., as a means to uncover how galaxies evolve over time in different environments.  \\

\subsection{Using \hi to probe gas inflow and removal} 

Galaxies form and evolve via the acquisition of gas from the
intergalactic medium (IGM).  In star-forming (blue cloud) galaxies,
gas infall is expected to occur over a Hubble time to keep them
forming stars at a relatively high rate
(\cite{Fraternali&Tomassetti12}).  Indirect evidence for gas infall
comes from estimates of gas depletion times (\cite{Bigiel+11}), the
reconstruction of star formation histories (\cite{Panter+07}), and
chemical evolution models of the Milky Way
(\cite{Schoenrich&Binney09}.  However, at present the observational
evidence for gas inflow into galaxies is rather scant, and it is not
clear how and in what form it takes place
(\cite{2008A&ARv..15..189S}).

\begin{figure}[h]
\centering
\includegraphics[width=15cm]{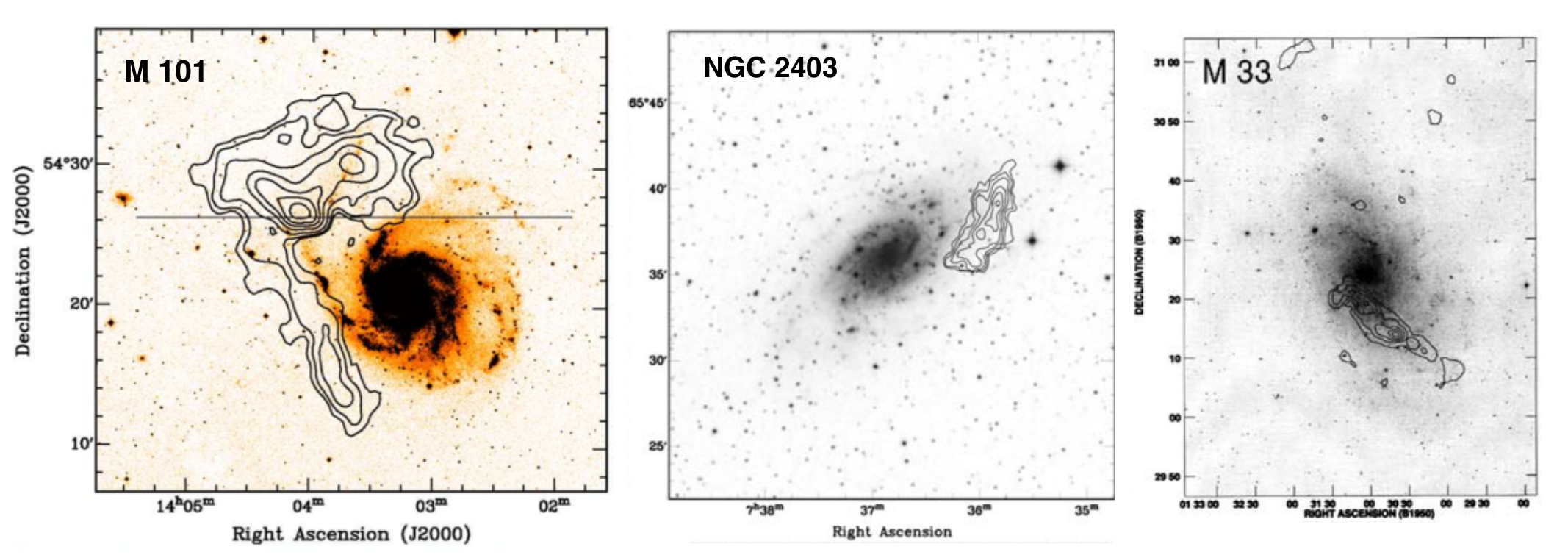}
\caption{\small
Three examples of possible accretion of \hi\ in nearby spirals indicated by the presence of gas complexes at anomalous velocities: M~101 (\cite{1988AJ.....95.1354V}), NGC~2403 (\cite{2002AJ....123.3124F}), and M~33 (\cite{2008A&ARv..15..189S}). 
}
\label{figure_accretion}
\end{figure}

Inflow of gas into galaxies has a long history dating back to the
discovery of the high-velocity clouds (HVCs) in the Milky Way
(\cite{Wakker&vanWoerden97}).  To date, the origin of these clouds
remain a mystery, and their contribution to the feeding of the
Galactic star formation is uncertain (\cite{Putman+12}).  Deep HI
observations of external galaxies also show similar features in HI
(\cite{Westmeier+05}).  Fig.\ \ref{figure_accretion} shows examples of
massive clouds/filaments observed in projection over the bright
optical disk of three nearby galaxies. These filaments/complexes are
well separated in velocity from the \hi\ in their disks. These
structures have \hi\ masses similar to that of the Galactic HVC
Complex C.  It is unlikely that they result from a galactic fountain
and they may be the signs of either minor mergers or the infall of
cosmological filaments (\cite{Fernandez+12}).

The typical sensitivities achieved with the current interferometers
for deep HI observations like those of NGC\,2403
(\cite{2002AJ....123.3124F}) are of order $10^{19} {\rm cm^{-2}}$.
SKA will improve this sensitivity by $2$ orders of magnitude, keeping
the spatial resolution very high.  This will have a tremendous impact
on our understanding of gas inflow, as we will finally be able to
trace the origin of these structures and detect gas falling into
galaxies from the IGM.  At present, the only information about low
column density material around galaxies comes from absoption line
studies towards AGNs.  From the study of Ly$\alpha$ absorbers, it is
clear that there are vast reservoirs of {\it cold} gas surrounding
both star-forming and early type galaxies (\cite{Tumlison+13}).
However their origin, morphology, and fate are not clear; for instance
whether they are gas clouds falling into the galaxies or simply
floating in the galactic coronae remains ambiguous.  This can be
explored with the SKA.

Galaxies also eject gas from their disks as a consequence of stellar
winds and supernova explosions.  The ejected gas is likely multiphase
but a large fraction of it is likely cold (\cite{Melioli+09}) or will
cool rapidly (\cite{Houck&Bregman90}).  Observations show HI at large
distance from the Milky Way's disk associated with superbubble
expansion (\cite{Pidopryhora+07}) and gas at high velocities escaping
from holes in the disks of external galaxies (\cite{Boomsma+08}).

Such HI outflows produce large layers of extra-planar gas that contain
typically 10\% of the HI mass of a galaxy, reaching up to 25\% in
exceptional cases (\cite{2007AJ....134.1019O,2008A&ARv..15..189S}).
The study of the kinematics of \HI\ halos gives essential information
about the exchange of material and angular momentum between galactic
disks and the IGM (\cite{Marinacci+10}).  However, at the moment, the
overall picture is not understood, as these low column density
features are not seen around all galaxies and in several cases are not
very extended (\cite{2011A&A...526A.118H}).  Going down to a
sensitivity well below $\times 10^{18}\, {\rm cm^{-2}}$ will provide
unprecented information about gas escaping from galaxies and how this
interacts with the surrounding medium.  With SKA2 we will eventually
be able to fully trace the gas cycle from disks to the IGM and
understand its crucial role for the evolution of star-forming
galaxies.

\begin{figure}[t]
\centering
\includegraphics[width=15cm]{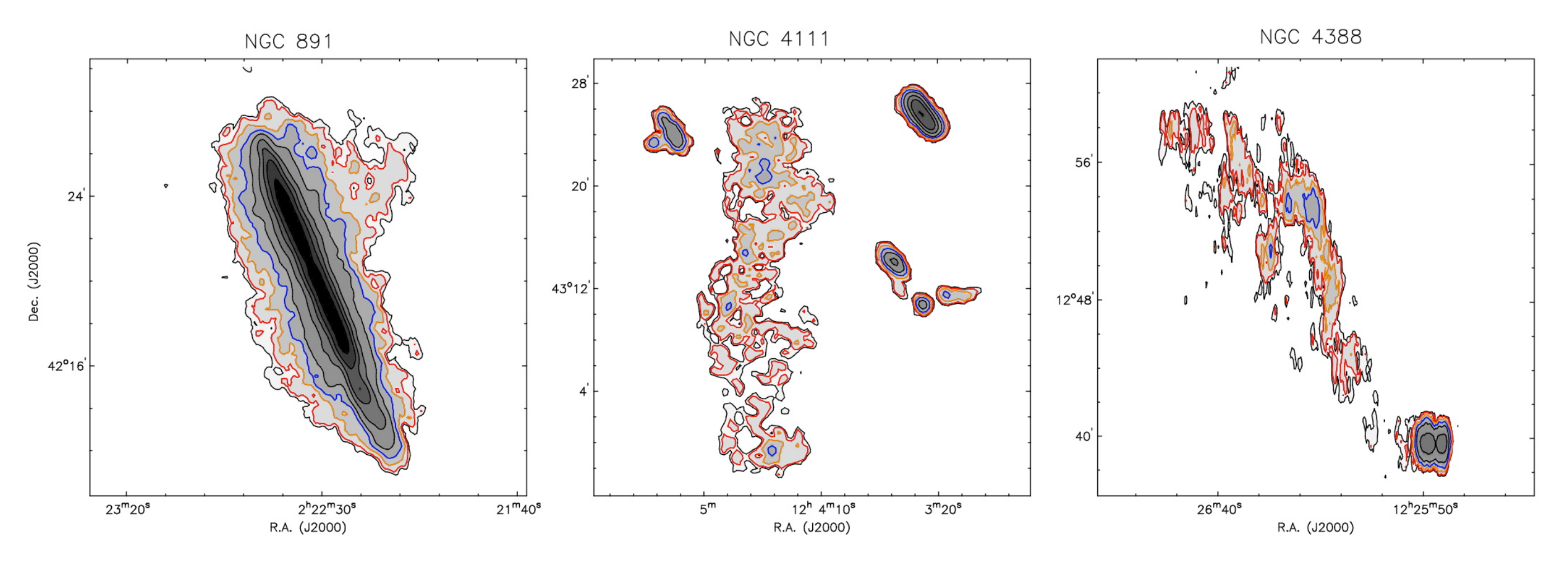}
\caption{\small
Three examples of galaxies undergoing gas accretion and gas removal events. From left to right: NGC891 (\cite{2007AJ....134.1019O}), NGC4111 (\cite{2004ogci.conf..394V}), and NGC4388 (\cite{2005A&A...437L..19O}). The lowest contours levels are $2 \times 10^{20} {\rm cm}^{-2}$ (red), $5 \times 10^{20} {\rm cm}^{-2}$ (orange) and $10 \times 10^{20} {\rm cm}^{-2}$ (blue). It is clear that the signs of gas accretion (NGC891, NGC4111) and of gas stripping (NGC4388) require column density sensitivities of well below $10^{21} {\rm cm}^{-2}$. 
}
\label{limiting_NHI}
\end{figure}

Gas can also be removed from galaxies via tidal interactions
(\cite{2008A&ARv..15..189S}) and ram pressure stripping in galaxy
clusters (\cite{Kenney+04}, see also Fig.\ \ref{limiting_NHI}).  At
the moment, these observations are limited to very few cases of mostly
very massive structures (\cite{2005A&A...437L..19O}) although \hi\
deficiencies are also found in small but dense environments, such as
Hickson compact groups (\cite{2001A&A...377..812V}).  SKA will change
this dramatically, giving the possibility to study tidal interactions
and minor mergers in unprecentedly large galaxy samples.  We will be
able to investigate the prevalence of these phenomena, the accretion
and removal rates from interaction and how all these properties change
as a function of environment and time.  The study of gas removed from
galaxies in dense environments will also finally solve the problem of
star formation quenching, something that is considered of fundamental
importance for the migration of star-forming galaxies towards the red
sequence (\cite{Blanton&Moustakas09}).

\subsection{The TF relation} 

The Tully-Fisher relation (TFr), originally describing a correlation
between a galaxy's intrinsic luminosity and the distance-independent
width of its global \hi{} profile (\cite{1977A&A....54..661T}), is the
most important observed scaling relation for gas-rich, late-type
galaxies.  Although the TFr is empirical in nature, it seems
astrophysically plausible that a galaxy's luminosity, as a proxy for
its stellar mass, should be related to its gravitational potential as
traced by the rotation speed of its gas disk, observed through the
Doppler-broadened width of its \hi{} emission line. The presence of
dark matter and the detailed distribution of the visible mass inside
galaxies complicate this naive interpretation
(\cite{1995MNRAS.273L..35Z}). Nevertheless, the TFr serves as a
valuable tool to obtain distances to galaxies in the local universe
where peculiar velocities dominate over the Hubble flow
(\cite{2012ApJ...749...78T}), while its statistical properties
constrain semi-analytical and hydrodynamical simulations of galaxy
formation and evolution.  Studies of the detailed \hi{} kinematics of
regularly-rotating galaxies for which the extended \hi{} gas disk is a
proper tracer of the gravitational potential, and for which good
estimates of their relative stellar masses are available, indicate
that the TFr is fundamentally a correlation between the maximum
rotational velocity of the dark matter halo and the embedded total
baryonic mass, regardless of how this baryonic mass is distributed
(\cite{2000ApJ...533L..99M,2001ApJ...563..694V,2005ApJ...632..859M}).

Major advances in photometric imaging technologies have significantly
improved our measures and understanding of a galaxy's luminosity to an
extent that, combined with spectroscopic information, it has become
common practice to replace a galaxy's luminosity in the TFr by its
derived stellar mass, or even its baryonic mass when taking the atomic
and estimated molecular gas masses into account
(\cite{2005ApJ...632..859M,2014AJ....147..134Z}). Improving our
understanding of the width of the global \hi{} line as a kinematic
measure has received little attention so far, and even the most recent
studies still use this quantity
(\cite{2013ApJ...765...94S,2014AJ....147..134Z}). As galaxies are
usually not spatially resolved by single-dish radio telescopes, the
shape and width of the global \hi{} profile is the result of a
convolution of the detailed distribution of the \hi{} gas in a galaxy
with the detailed geometry and kinematics of the gas disk, including
possible warps, streaming motions, asymmetries and the overall shapes
of rotation curves, which may not always be monotonically rising to a
constant circular velocity
(\cite{1981AJ.....86.1791B,1981AJ.....86.1825B,1989A&A...223...47B,1991AJ....101.1231C,2001A&A...370..765V,2007MNRAS.376.1513N,2009A&A...493..871S}). Hence,
the width of the global \hi{} profile is only a proxy for the
rotational speed of a galaxy, while the latter may not even be
properly defined and identified for non-flat rotation curves, and
trends in rotation curve shapes along the TFr may introduce
significant biases in scatter and slope
(\cite{2001ApJ...563..694V,2007MNRAS.381.1463N}). Imaging the geometry
and kinematics of a galaxy's extended \hi{} disk with interferometers
is the only observational avenue to identify the relevant kinematic
measure for the TFr and to significantly improve our understanding of
the underlying astrophysics.

It is expected that upcoming \hi{} imaging surveys with ASKAP,
WSRT/APERTIF and MeerKAT will further reduce the observed scatter in
the TFr by spatially resolving the \hi{} kinematics thereby improving
TF-based distances to galaxies in the local universe. These imaging
surveys will also establish the statistical properties of the TFr at
modest redshifts ($z<0.2$) using spatially resolved \hi{} kinematics,
and up to $z<0.5$ using global \hi{} line widths. It is unlikely that
the SKA will significantly improve upon these studies. However, given
the anticipated sensitivity and angular resolution of the SKA, it will
be possible to exploit the TFr to study the evolution of galaxies over
cosmic time by investigating the slope, scatter and zero-point as a
function of redshift and cosmic environment, as well as global galaxy
properties such as morphological type and (specific) star formation
rate. This requires adequate angular resolution and column density
sensitivity to image the \hi{} kinematics in the outer gas disks of
galaxies at higher redshifts, as well as survey volumes that encompass
all cosmic environments with sufficient statistical significance to
mitigate cosmic variance. In order to assess the kinematical state of
the largest \hi{} disks in M$^*_{HI}$ galaxies at a redshift of
$z=0.5$, a spatial resolution of 10 kpc would be sufficient,
corresponding to an angular resolution of 1.6 arcseconds at a
frequency of 950 MHz. Based on studies of nearby galaxies, the
required column density sensitivity at this angular resolution would
need to be $\sim 5\times 10^{20}$ [cm$^{-2}$]. This could be
achievable with an ultra-deep survey using SKA1-MID but seems out of
reach for SKA-SUR. The volume surveyed by a single SKA1-MID pointing
at $z=0.5$ at 950 MHz is large enough to sample all cosmic
environments with a sufficient number of detections to provide
adequate statistics on the TFr. Pushing TFr studies based on resolved
\hi{} kinematics to even higher redshifts would require the power of
SKA2, in particular given the required column density sensitivities.

\begin{figure}[t]
\centering
\includegraphics[width=15cm]{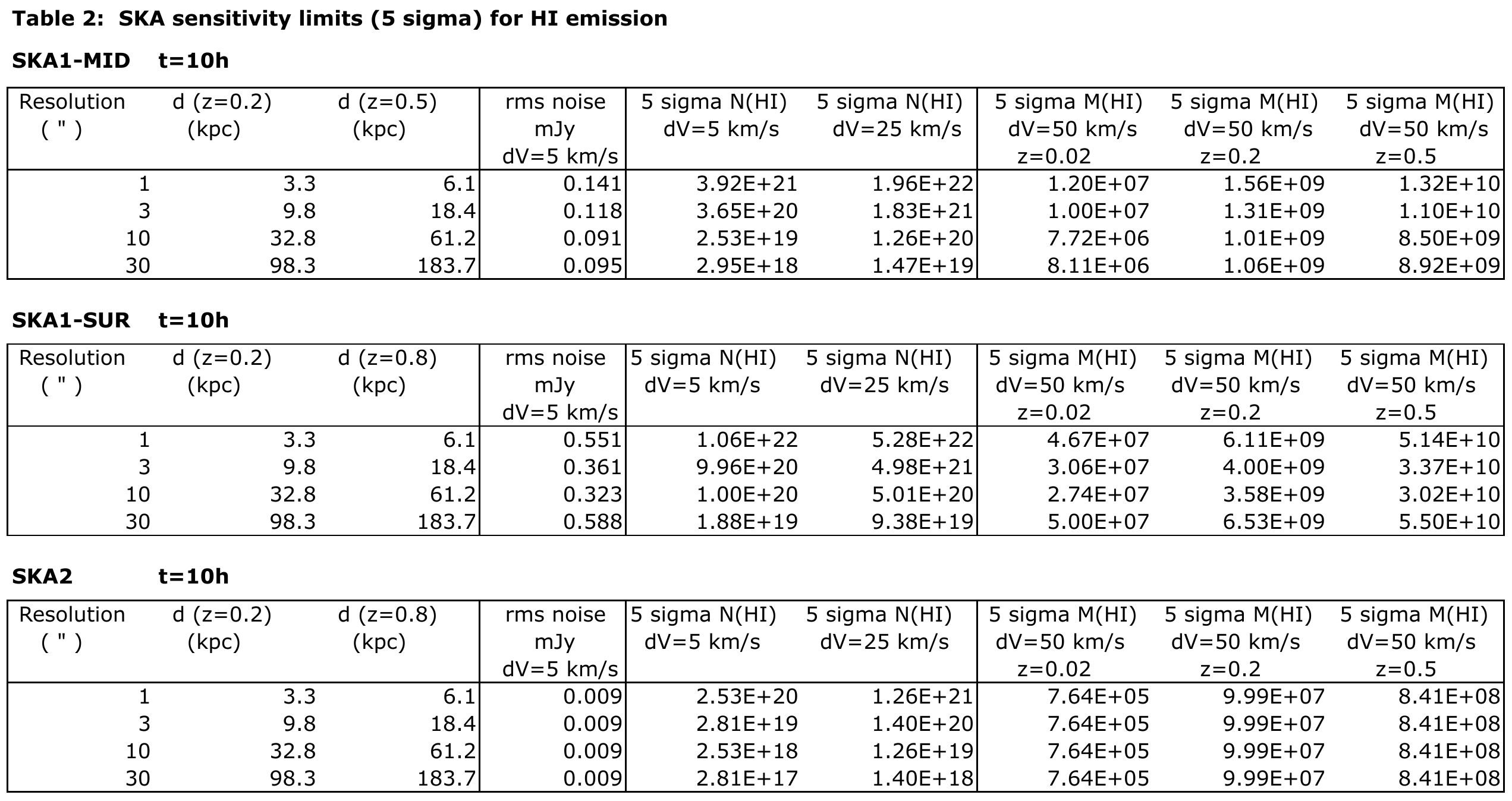}
\end{figure}

\section{SKA HI Strawman Surveys}

\subsection{Survey scenarios and requirements}

The science goals outlined in section 3 should lead us to define the
most efficient observing strategies for obtaining the data, given the
the technical capabilities of the SKA, in the first phase and eventual
final form. A review of the science addressed in section 3 makes clear
immediately that it is of the greatest importance not only to have a
sufficiently large number of objects for statistical studies probing
different regions of the universe, both spatially and in cosmic time,
but also to have the appropriate linear resolution and column density
sensitivity to probe the phenomena of interest. For studies involving
the \hi{} mass function, it is important to push the detection limit
to objects with small \hi{} masses ($< 10^{7}$ \msol) without having
to resolve the \hi{} distributions. For all other science goals it is
extremely important to resolve the galaxies and be able to detect
\hi{} column densities to levels well below \NHI\ $\sim 10^{21} {\rm
  cm}^{-2}$. This is beautifully illustrated in Fig.\
\ref{limiting_NHI}.

Using the specifications of the SKA baseline design (\cite{SKA_BD}),
one can derive the observing parameters required to reach a depth of
\NHI\ $\sim 10^{20} {\rm cm}^{-2}$ at reasonable resolution. Table 2
provides basic input to these considerations. The numbers are based on
the latest sensitivity estimates for SKA1-MID, SKA1-SUR and
SKA2. Table 2 provides information for four different resolutions
because \NHI\ sensitivity scales with resolution. The corresponding
linear resolution is given for two nominal distances. In addition,
Table 2 gives the \hi\ mass limits for a 10 hour integration assuming
a profile width of 50 km s$^{-1}$ for three nominal distances. For
reference, the canonical value for $M_{\rm HI}^*$ is $\sim 8 \times
10^{9}$ \msol (\cite{2003AJ....125.2842Z, 2005MNRAS.359L..30Z,
  2011AJ....142..170H}).

\begin{figure}[t]
\centering
\includegraphics[width=15cm]{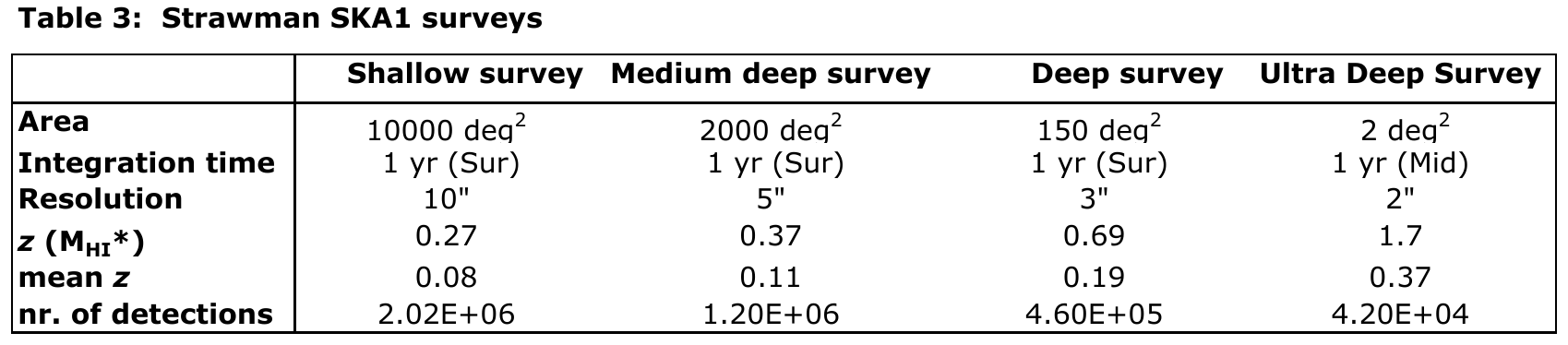}
\end{figure}

\begin{figure}[t]
\centering
\includegraphics[width=10cm]{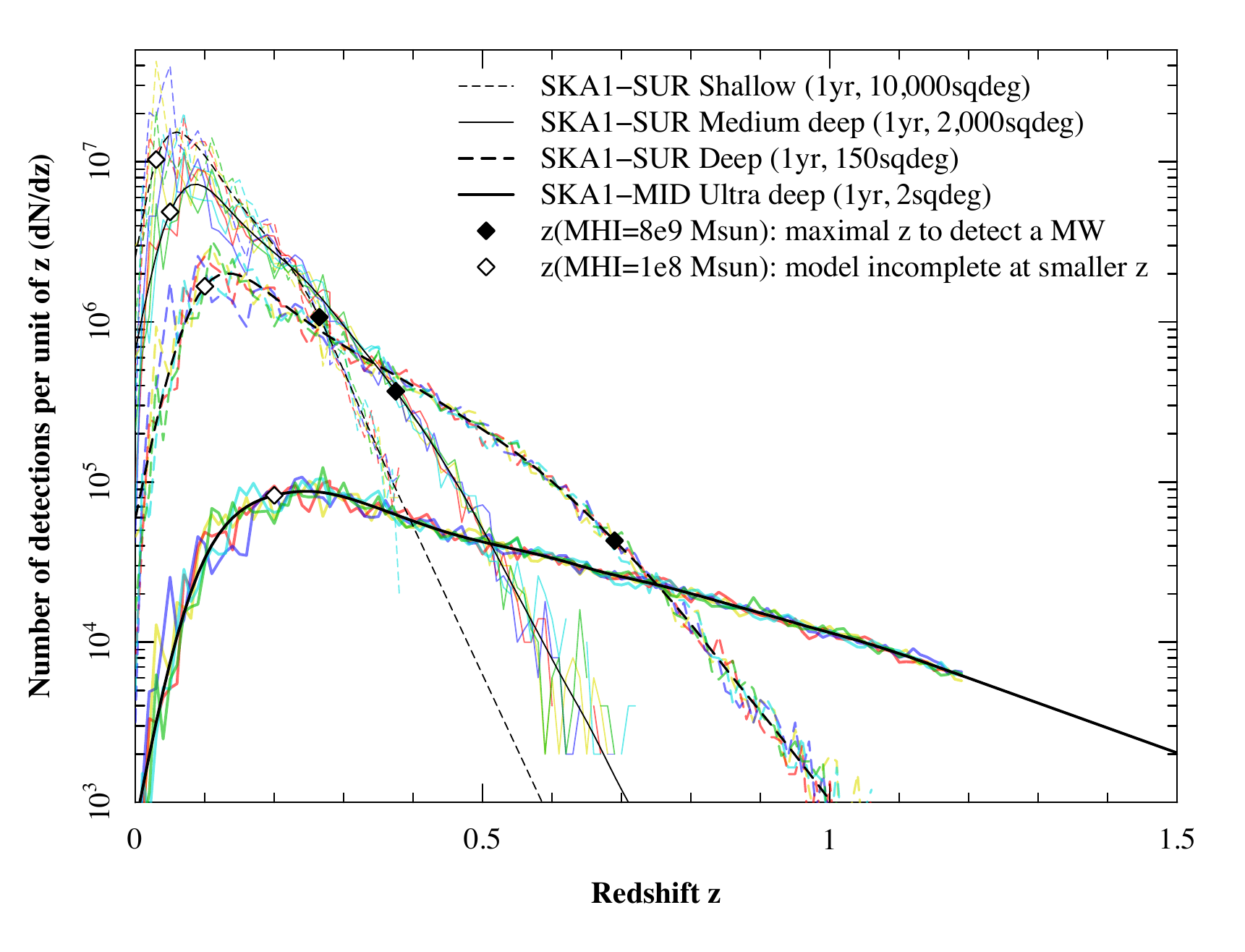}
\caption{Distribution over redshift of the four surveys given in Table 3 based on the S3-SAX simulation (\cite{2009ApJ...703.1890O,2009ApJ...698.1467O})}\label{dNdzplot}
\end{figure}

It is clear that in order to design the right surveys for the science
described in this chapter, trade-offs need to be made between
resolution, integration time, and area covered. Table 3 provides an
attempt to do this for SKA1. Survey areas are chosen such that the
total integration time remains realistic, while the appropriate
resolution and column density are chosen to match the science goals
outlined in the previous sections. For example, a survey of the entire
southern sky to a depth of \NHI\ $\sim 10^{20}$ at a resolution of 2"
would take half a century with SKA1, so a better approach is to cover
a smaller area at lower but still useful resolution, leaving
substantial improvement on a large \hi\ survey such as WALLABY to
SKA2. Which surveys cover substantial gound in redshift space is very
well illustrated in Fig.~\ref{dNdzplot}, showing the distribution over
redshift of the four surveys from Table 3. The distributions represent
five statistically independent realisations of a 100 deg$^2$ sky
model, truncated to $z=1.2$, based on the S3-SAX simulation
(\cite{2009ApJ...703.1890O,2009ApJ...698.1467O}) using the telescope
simulations of \cite{Popping2014} which are consistent with the
sensitivity calculations used here. In other words SKA1 is the
appropriate instrument for addressing follow-up science from the SKA
pathfinders; SKA2 provides a major step towards pushing our observing
capabilities to higher redshifts, i.e., look back times. In order to
obtain a good census of galaxy evolution and the influence of the
environment over cosmic time, one has to design surveys in such a way
that SKA1 probes the epochs $0.1<z<0.5$ with similar linear resolution
and \NHI\ sensitivity as APERTIF, ASKAP and MeerKAT will probe at
$0.02<z<0.2$, while SKA2 will eventually provide the data required to
get the full picture of \hi\ evolution from the current epoch to
$z\sim2$ when the star formation rate density in the evolving universe
is at its peak.

\section{Summary and conclusions}

Our current \hi{} view of the universe is restricted to the Local
Volume and hence only informs us about gas-related processes in
galaxies in a late state of their cosmic evolution.  Semi-analytical
and hydrodynamical models, however, suggest significant changes in the
\hi{} properties of galaxies beyond the Local Universe.  We are poised
to start testing these models with upcoming surveys on the SKA
pathfinder instruments over the next few years.  These facilities will
provide much improved statistics on the \hi{} properties of galaxies
and will begin to open a window on the \hi{} universe beyond the Local
Volume. The SKA1 (and eventually SKA2) will truly revolutionise
studies of \hi{} in galaxies by enabling observations at higher
angular resolutions and with sensitivities that are an order of
magnitude greater than currently possible.  Most importantly, the SKA1
will allow us to reach out to distances and lookback times spanning
more than half the age of the universe.  Combined with data at other
wavelengths and detailed theoretical models, \hi{} observations with
the SKA1 will enable us to trace the baryon cycle in galaxies from the
IGM and back out again in different environments.  The SKA2 offers the
exciting prospect of directly measuring the HI masses of the nearest
Lyman limit systems and performing precision cosmology with the
billion galaxy survey.  Ultimately, we will reveal the evolution of
the distribution and kinematics of gas in galaxies back to $z\sim2$
when the cosmic star formation rate density was at its peak and an
order of magnitude higher than the present, thereby building a
comprehensive picture of galaxy evolution over cosmic time.

\section*{Acknowledgments}

We thank various colleagues for valuable input to this chapter. JMvdH
acknowledges support from the European Research Council under the
European Union's Seventh Framework Programme (FP/2007-2013) / ERC
Grant Agreement nr. 291531. BC is the recipient of an Australian
Research Council Future Fellowship (FT120100660). MPH is supported by
US-NSF/AST-1107390 and the Brinson Foundation. LVM has been supported
by grants AYA2011-30491-C02-0S1 and TIC-114.

 
%
%

\bibliographystyle{apj}

\end{document}